\documentclass[aps,superscriptaddress,showpacs,nofootinbib,eqsecnum]{revtex4}

\begin{document}

\title{Abelian Higgs Model Effective Potential in the Presence of Vortices}

\author{Rudnei O. Ramos}
\email{rudnei@uerj.br}
\affiliation{Departamento de F\'{\i}sica Te\'orica,
Universidade do Estado do Rio de Janeiro, 20550-013
Rio de Janeiro, RJ, Brazil}

\author{J. F. Medeiros Neto}
\email{jfmn@ufpa.br}
\affiliation{Departamento de F\'{\i}sica Te\'orica,
Universidade do Estado do Rio de Janeiro, 20550-013
Rio de Janeiro, RJ, Brazil}

\affiliation{Instituto de F\'{\i}sica,
Universidade Federal do Par\'a, 66075-110 Belem, Par\'a, Brazil}

\author{Daniel G. Barci}\thanks{Regular Associate of
the Abdus Salam International Centre for Theoretical Physics (ICTP),
Trieste, Italy}
\email{barci@uerj.br}
\affiliation{Departamento de F\'{\i}sica Te\'orica,
Universidade do Estado do Rio de Janeiro, 20550-013
Rio de Janeiro, RJ, Brazil}

\author{Cesar A. Linhares}
\email{linhares@dft.if.uerj.br}
\affiliation{Departamento de F\'{\i}sica Te\'orica,
Universidade do Estado do Rio de Janeiro, 20550-013
Rio de Janeiro, RJ, Brazil}

\begin{abstract}

We determine the contribution of nontrivial vacuum (topological)
excitations, more specifically vortex--strings of the Abelian Higgs model
in $3+1$ dimensions, to the functional partition function. By expressing
the original action in terms of dual transformed fields we make explicit
in the equivalent action the contribution of the vortex--strings
excitations of the model. The effective potential of an appropriately
defined local vacuum expectation value of the vortex--string field in the
dual transformed action is then evaluated both at zero and finite
temperatures and its properties discussed in the context of the finite
temperature phase transition.

\end{abstract}


\maketitle

\section{Introduction}

The study of phase transitions in quantum field theory has a long
history, since the first works on the subject \cite{KL, dolan, weinb}
(see also the Refs. \cite{kapusta, bellac}) and it is still a highly
active area of research motivated by several open problems in QCD phase
transitions, grand-unified theory phase transitions and many other
subject areas including also condensed matter physics problems
\cite{lubensky}. One basic mechanism we are usually interested in these
studies is how the variation of an external quantity like temperature,
density or external fields may act and change different physical
quantities in a given system or the study, for instance, of how
symmetries may change under the variation of temperature, like in
symmetry breaking phase transitions. One very common and extremely
useful tool in the latter problem is the use of effective potentials for
appropriate order parameters characterizing the possible phases of the
system (at equilibrium), like the vacuum expectation value of a Higgs
field in gauge field theories, determined as some constant (in space and
time) solution of the effective field equations.

Around the same time of these studies on symmetry breaking/restoring
phase transitions on gauge field theories, it was also realized that
symmetry breaking in gauge field theories could give rise to nontrivial
and nonperturbative stable solutions of the field equations of motion.
This is the case, for example, of the magnetic vortex solutions in a
$U(1)$ symmetry broken Abelian gauge field theory \cite{nielsen-olesen}
or magnetic monopoles in $O(3)$ or $SU(2)$ symmetry broken non-Abelian
gauge field theories \cite{hooft, polyakov}, which are only a few
examples among several other topological-like nontrivial vacuum field
solutions that have been exhaustively studied to date (for reviews, see
for instance Refs. \cite{coleman, rajaraman}). Extra interest on these
field solutions is also due to the fact that, since these nonlocal
vacuum structures are expected to emerge in most of the grand unified
phase transitions in the early universe, they may have important
cosmological consequences (for a detailed account see e.g. Ref.
\cite{review}).

In the present paper we consider the case of phase transition in the
Abelian Higgs model from the viewpoint in which the phase transition at
finite temperatures is driven by a condensation of magnetic vortices.
This is not an entirely novelty in the sense that there are a lot of
examples in which phase transitions are driven by topological defects in
quantum field theory as well as in condensed-matter physics
\cite{kleinertI-II}. In fact, it has long been believed that, close to
the critical point, the condensation of inhomogeneous configurations,
solutions of the field equations, is able to provide a much better
description of the phase transition as compared to mean field methods,
e.g., using the sole contribution of constant, homogeneous field
configurations in the partition function, as it is the case of the
standard derivations of the finite-temperature effective potential in
field theories. {}For instance, topological configurations, like strings
in the Abelian Higgs model, have previously been studied in this context
of phase transitions by computing the free energy associated to these
configurations, e.g., by semiclassically expanding the quantum fields
around the vortex-string {\it classical} solution
\cite{freeenergy,copeland}. The problem with this approach of
considering the contribution of topological configurations to the
effective action in a semiclassical way is the intrinsic difficulty of
computing the effective action, which becomes highly nonlocal, so only
the first order loop terms can be computed analytically and to go beyond
numerical methods have to be employed. An alternative approach to the
semiclassical one that also has been used is directly quantizing the
topological excitations and representing them as (nonlocal) quantum
fields (see for instance the approach of Refs. \cite{marino} and
references therein). But this is also problematic since we are only able
to compute lowest-order correlation functions of the quantal topological
field and even so, the still nonlocal character of these functions
besets a simple derivation. To circumvent these problems, in this paper
we adopt an alternative intermediate derivation between the latter two,
which make use of the concept of duality \cite{fisher}. Using this
technique it is possible to conveniently rewrite the original action for
the Goldstone modes of the broken symmetry, in terms of a dual action
describing the topological defect currents and its interactions mediated
by a dual {\it antisymmetric} tensor field.

We here consider the finite-temperature version of the Abelian Higgs
model, which is then treated along a formalism developed long ago by the
authors of Refs. \cite{suga1, seo-suga, kawai}. In this formalism, a
dual transformation is applied to the Higgs model partition function in
order to show the contributions from topological excitations in a more
explicit manner. An antisymmetric tensor auxiliary field is introduced
and, after functional integration of the original electromagnetic vector
field, the action of this dual model assumes the form of a relativistic
hydrodynamics in the sense of Kalb--Ramond \cite{kalbramond} and Nambu
\cite{nambu0, nambu}. The formalism may be generalized to non-Abelian
gauge fields \cite{okawa}. In more recent years this formalism has been
generalized to extended objects in higher dimensions (D-branes) in
string theory \cite{suga2}. Also, in another kind of application, this
duality approach has been used in the study of vortices in superfluidity
models \cite{superfluid}.

The next step in this mechanism, is to rewrite the sum over all possible
distributions of the topological number density which appears in the
partition function as a functional integration over some functional
fields. This procedure was introduced previously in $U(1)$ lattice gauge
theory \cite{stone} and later used in the Abelian Higgs model by several
authors \cite{bardacki, kawai}. In this paper we use these techniques to
calculate the contribution of the topological defects in the Abelian
Higgs model to the one-loop effective potential, which can now be
expressed directly in terms of the expectation value of a quantum vortex
field. {}From this effective potential we have calculated the vortex
condensation temperature obtaining a result compatible with previous
estimations based on the statistical distribution of classical strings
\cite{copeland}. We have also checked that this temperature is different
than the usual mean field critical temperature of the model when the
inhomogeneous topological field configurations are neglected. This then
makes possible to access in an analytical way the importance of these
topological configurations during a phase transition.

The remaining of this paper is organized as follows. In Sec. \ref{model}
we introduce the model. In Sec. \ref{dual} we calculate the dual action
showing how the topological defects explicitly show up in this
formalism. We discuss the issue of gauge invariance and the equivalence
between the original and the dual model at the effective potential
level. In Sec. \ref{effpot} we
calculate the contribution of the topological defects to the effective
potential and evaluate the condensation temperature. Our final
considerations and conclusions are given in Sec. \ref{conclusions}. An
appendix is included to review and show some of the technical details of
the formalism we have used here.

\section{The model}
\label{model}

The model we consider is the Abelian Higgs model with Lagrangian
density for a complex scalar field $\phi$ and gauge field $A_\mu$,

\begin{equation}
{\cal L} = - \frac{1}{4} F_{\mu \nu } F^{\mu \nu }+|D_\mu \phi|^2
-V(\phi)\;,
\label{lagr}
\end{equation}

\noindent
where, in the usual notation,
${}F_{\mu \nu } =\partial _\mu A_\nu -\partial _\nu A_\mu$,
$D_\mu = \partial _\mu -ieA_\mu$ and $V(\phi)$ is a symmetry
breaking potential given by

\begin{equation}
V(\phi) = - m_{\phi }^2\left| \phi \right| ^2+\frac \lambda
{3!}\left( \left| \phi \right| ^2\right) ^2 \;.
\label{pot}
\end{equation}

\noindent
The symmetry breaking $U(1) \to 1$ with homotopy group $\pi_1 \neq 1$
indicates the existence of string-like topological excitations in the
system (for an extended introduction and review see e.g. Ref.
\cite{review}). {}For example, for a unit winding string solution
along the $z$ axis, the classical field equations of motion obtained
from the Lagrangian density (\ref{lagr}) admit
a stable finite energy configuration describing the string and given
by (using the cylindrical coordinates $r,\theta,z$)

\begin{eqnarray}
\phi_{\rm string} &=& \frac{\rho(r)}{\sqrt{2}} e^{i \theta}\;,
\label{phi string}
\\
A_{\mu, {\rm string}}  &=& \frac{1}{e} A(r)\; \partial_\mu \theta\;,
\label{A string}
\end{eqnarray}

\noindent
where the functions $\rho(r)$ and $A(r)$ vanish at the origin and have
the asymptotic behavior $\phi(r \to \infty) \to \rho_v \equiv \sqrt{6
m_\phi^2/\lambda}$ and $A(r \to \infty) \to 1$. The functions $\rho(r)$
and $A(r)$ are obtained (numerically) by solving the classical field
equations. If we write the field $\phi$ as $\phi =\rho \exp (i\chi
)/\sqrt{2}$, then from (\ref{phi string}) and (\ref{A string}) for the
string, at spatial infinity $\rho$ goes to the vacuum $\rho_v$ and
$A_\mu$ becomes a pure gauge. This also gives, in order to get a finite
energy for the string configuration, that $\partial_\mu \chi = e A_\mu $
at $r\to \infty$, so $D_\mu \phi=0$. This leads then that, by taking
some contour $C$ surrounding the symmetry axis, and using Stokes'
theorem, to the nonvanishing magnetic flux

\begin{equation}
\Phi = \oint A_\mu dx^\mu = \oint \partial_\mu \chi dx^\mu = 2\pi/e\;.
\label{flux}
\end{equation}

\noindent
Since $\phi$ must be single-valued, the Eq. (\ref{flux}) implies that on
the string $\chi$ must be singular. Therefore, the phase $\chi$ can be
separated into two parts: in a regular part and in a singular one, 
due to the string
configuration. We will use this latter fact in the next section when
describing the topological vortex string contributions to the partition
function, which are then characterized by multivalued (or singular)
phases of the scalar field.

\section{The Dual-transformed action}
\label{dual}

Let us start by writing the partition function for the Abelian Higgs
model (\ref{lagr}), which, in Euclidean space-time is given by

\begin{equation}
Z[\beta ]=\int \mathcal{D}A\mathcal{D}\phi \mathcal{D}\phi^*
\exp\left\{ -S\left[ A_\mu ,\phi ,\phi^*\right] - S_{GF} \right\}  \;,
\label{ZAphi}
\end{equation}

\noindent
where in the above expression $S$ denotes the Euclidean action,

\begin{equation}
S \left[ A_\mu ,\phi ,\phi^*\right] =\int_0^\beta d\tau \int
d^3x\left[ \frac{1}{4} {}F_{\mu \nu }{}F_{\mu \nu }+|D_\mu \phi|^2
+V(\phi)\right] \;,
\label{actini}
\end{equation}

\noindent
where $\beta=1/T$ is the inverse of the temperature and $S_{GF}$ in
(\ref{ZAphi}) is some appropriate gauge-fixing and ghost term that must
be added to the action to perform the functional integral over the
relevant degrees of freedom. We will come back later to this term and
explicitly fix it within the formalism described below. Note also that
the functional integral in Eq. (\ref{ZAphi}) is to be performed over the
bosonic scalar and gauge fields satisfying the usual periodic boundary
conditions in imaginary time with period $\beta =1/T$
\cite{kapusta,bellac}.

By writing the complex Higgs field $\phi$ in the polar parameterization
form $\phi =\rho e^{i\chi }/\sqrt{2}$, the functional integration
measure in Eq. (\ref{ZAphi}) is changed to

\begin{eqnarray}
\mathcal{D}\phi \mathcal{D}\phi^*\to \mathcal{D}\rho \mathcal{D}
\chi \left( \prod_x\rho \right) \;,
\end{eqnarray}

\noindent
and the quantum partition function becomes

\begin{equation}
Z=\int \mathcal{D}A_\mu \mathcal{D}\rho \mathcal{D}\chi \left( \prod_x\rho
\right) \exp \left\{ -S\left[ A_\mu ,\rho ,\chi \right] -S_{GF}\right\} \;,
\label{ZArhochi}
\end{equation}
with

\begin{equation}
S\left[ A_\mu ,\rho ,\chi \right] =\int d\tau d^3x\left[ \frac 14F_{\mu \nu
}F_{\mu \nu }+\frac 12\left( \partial _\mu \rho \right) ^2+\frac 12\rho
^2\left( \partial _\mu \chi +eA_\mu \right) ^2-\frac{m_\phi ^2}2\rho
^2+\frac \lambda {4!}\rho ^4\right] \;.
\label{SArhochi}
\end{equation}

In order to make explicit the contribution of the nontrivial
topological field configuration in the partition function
(\ref{ZArhochi}), it is more convenient to work with the dual version of
Eq. (\ref{SArhochi}). To achieve this equivalent dual action we start by
splitting the scalar phase field $\chi $ in its regular and singular
terms, $\chi =\chi _{\mathrm{reg}}+\chi _{\mathrm{sing}}$. Lets for now,
for convenience, omit the gauge fixing term $S_{GF}$ in Eq. (\ref{ZArhochi}) 
and re-introduce
it again in the final transformed action. {}Following
e.g. the procedure of Refs. \cite{klee,orland,chernodub,antonov1,kleinert}, 
the functional
integral over $\chi $ in Eq. (\ref{ZArhochi}) can then be rewritten as

\begin{eqnarray}
\lefteqn{\int \mathcal{D}\chi \,\exp \left[ -\int d^4x\frac 12\rho ^2\left(
\partial _\mu \chi +eA_\mu \right) ^2\right] }  \nonumber \\
&=&\int \mathcal{D}\chi _{\mathrm{sing}}\,\mathcal{D}\chi _{\mathrm{reg}}
\mathcal{D}C_\mu \left( \prod_x\rho ^{-4}\right) \,\exp \left\{ -\int
d^4x\left[ \frac 1{2\rho ^2}C_\mu ^2-iC_\mu \left( \partial _\mu
\chi _{\mathrm{reg}}\right) -iC_\mu \left( \partial _\mu \chi _{\mathrm{sing}
}+eA_\mu \right) \right] \right\}  \nonumber \\
&=&\int \mathcal{D}\chi _{\mathrm{sing}}\left( \prod_x\rho ^{-4}\right)
\mathcal{D}W_{\mu \nu }\,\exp \left\{ -\int d^4x\left[ \frac{\kappa ^2}{
2\rho ^2}V_\mu ^2+e\kappa A_\mu V_\mu +i\pi \kappa W_{\mu \nu }\omega _{\mu
\nu }\right] \right\} \;,
\label{dual2}
\end{eqnarray}

\noindent where we have performed the functional integral over
$\chi _{\mathrm{reg}}$ in the second line of Eq. (\ref{dual2}).
This gives a constraint on the functional integral measure,
$\delta (\partial _\mu C_\mu )$, which can be represented in a
unique way by expressing the $C_\mu $ in terms of an antisymmetric
field, $C_\mu = -i\frac \kappa 2\epsilon _{\mu \nu \lambda \rho
}\partial _\nu W_{\lambda \rho }\equiv \kappa V_\mu $, which then
leads to the last expression in Eq. (\ref{dual2}). $\kappa $ is
some arbitrary parameter with mass dimension and $\omega _{\mu \nu
}$ is the vorticity given only in terms of the singular phase part of
$\chi$,

\begin{eqnarray}
\omega _{\mu \nu }\equiv \frac 1{4\pi }\epsilon _{\mu \nu \lambda \rho
}\left( \partial _\mu \partial _\nu -\partial _\nu \partial _\mu \right)
\chi (x)\;.
\label{omega-munu}
\end{eqnarray}

\noindent
Next, in order to linearize the dependence on the gauge field in the
action we introduce a new antisymmetric tensor field $G_{\mu \nu }$
through the identity

\begin{equation}
\exp \left( -\frac 14\int d^4x {}F_{\mu \nu }^2\right) =\int \mathcal{D}G_{\mu
\nu }\,\exp \left[ \int d^4x\left( -\frac{\mu _W^2}4G_{\mu \nu }^2-\frac{\mu
_W}2\,\tilde{G}_{\mu \nu }F_{\mu \nu }\right) \right] \;,
\label{dual3}
\end{equation}
with
\begin{equation}
\tilde{G}_{\mu\nu} \equiv
\frac{1}{2}\epsilon_{\mu\nu\lambda\rho}G_{\lambda\rho}\;.
\end{equation}

\noindent
Substituting Eqs. (\ref{dual2}) and (\ref{dual3}) back into Eq.
(\ref{ZArhochi}),
we can immediately perform the functional integral over the $A_\mu $ field.
Taking also for convenience $e\kappa =\mu _W$, we then obtain for Eq.
(\ref{ZArhochi}) the result

\begin{eqnarray}
\lefteqn{Z= \int \mathcal{D}W_{\mu \nu }\mathcal{D}\chi _{\rm sing}
\mathcal{D}G_{\mu \nu }\delta \left[ \epsilon _{\mu \nu \alpha \beta
}\partial _\mu \left( G_{\alpha \beta }-\frac 12W_{\alpha \beta }\right)
\right] \,\mathcal{D}\rho \,\left( \prod_x\rho ^{-3}\right) }  \nonumber \\
 &&\times \exp \left\{ -\int d^4x\left[ \frac{\mu _W^2}4G_{\mu \nu }^2+
\frac{\mu _W^2}{2e^2\rho ^2}V_\mu ^2+\frac{1}{2}
\left( \partial_\mu \rho \right)^2-
\frac{m_\phi ^2}{2} \rho ^2+\frac{\lambda}{4!}\rho^4+
i\pi \frac{\mu _W}{e} W_{\mu \nu }\omega _{\mu \nu }\right] 
\right\}  \,.
\end{eqnarray}

\noindent
The constraint 
$\epsilon_{\mu \nu \alpha \beta }\partial_\mu 
\left( G_{\alpha \beta }-W_{\alpha \beta }\right) =0$ 
can be solved by setting

\begin{equation}
G_{\mu \nu }=W_{\mu \nu
}-\frac 1{\mu _W}\left( \partial _\mu B_\nu -\partial _\nu B_\mu \right),
\end{equation}

\noindent
where $B_\mu$ is an arbitrary gauge field, thus obtaining for the
partition function the  expression (and re-introducing the gauge 
fixing term)

\begin{equation}
Z=\int \mathcal{D}W_{\mu \nu
}\mathcal{D}\chi_{\mathrm{sing}}\,\mathcal{D}B_\mu\,\mathcal{D}
\rho \,\left( \prod_x\rho ^{-3}\right)\;\exp\left\{ -S_{\rm
dual}\left[ W_{\mu \nu },B_\mu, \rho ,\chi_{\rm sing} \right]
-S_{GF}\right\}, \label{ZWBrho}
\end{equation}
with

\begin{equation}
S_{\rm dual}= \int d^4x\left[ \frac{\mu _W^2}{2e^2\rho ^2}V_\mu
^2+\frac 14\left( \mu _WW_{\mu \nu }-\partial _\mu B_\nu +\partial
_\nu B_\mu \right) ^2+ \frac 12\left( \partial _\mu \rho \right)
^2-\frac{m_\phi ^2}2\rho ^2+\frac \lambda {4!}\rho ^4+i\pi
\frac{\mu _W}eW_{\mu \nu }\omega _{\mu \nu }\right] \;.
\label{SWBrho}
\end{equation}

\noindent This dual model is completely equivalent to the original
Abelian Higgs model in the polar representation given by Eqs.
(\ref{ZArhochi}) and (\ref{SArhochi}) and so, any calculations
done using (\ref{ZWBrho}) must lead to the same results as those
done with the original action. {}For example, if we compute the
effective potential for a constant scalar field configuration
$\rho_c$ from the latter should be the same as the one obtained by
the former. This we will check explicitly shortly. The advantage
of the dual version is that it explicitly exhibits the dependence
on the singular configuration of the Higgs field, making it
appropriate to study phase transitions driven by topological
defects. However, we need to be careful with gauge invariance, in
special in the dual model (\ref{SWBrho}), since it has more gauge
freedom than the original model. Now we come to the part
concerning the gauge fixing term $S_{GF}$ in (\ref{ZWBrho}).
{}From Eq. (\ref{SWBrho}) we see that the dual action exhibits
invariance under the double gauge transformation: the hypergauge
transformation

\begin{eqnarray}
\delta W_{\mu \nu }(x) &=&\partial_\mu \xi_\nu (x)-\partial_\nu \xi_\mu
(x)\;,  \nonumber \\
\delta B_\mu &=&\mu_W\xi_\mu (x)\;,
\label{gauge1}
\end{eqnarray}
and the usual gauge transformation

\begin{equation}
\delta B_\mu =\partial_\mu \theta (x)\;,  \label{gauge2}
\end{equation}

\noindent 
where $\xi_\mu (x)$ and $\theta (x)$ are arbitrary
vector and scalar functions, respectively. Choosing $\xi_\mu
=B_\mu $ in the first transformation is equivalent to fix the
gauge through the condition $B_\mu =0$ \cite{orland} and this is
equivalent to choose the unitary gauge in Eq. (\ref{ZWBrho}).

At this point, it would be interesting to analyze the gauge fixing
procedures for this model and to show that the resulting effective
potential does not depend on the gauge fixing parameters within our
parametrization choice for the complex scalar field. {}For simplicity,
we neglect at this time the last term in the exponential in Eq.
(\ref{SWBrho}) due to the vorticity. In order to evaluate the effective
potential we need to specify the gauge fixing term $S_{GF}$. To fix the
gauge for the antisymmetric tensor field, associated to the first gauge
transformation in Eq. (\ref{gauge1}), we need to introduce a vector
ghost field. We here do this in the same way the gauge is fixed and
corresponding ghost terms appear in the analogous case of choosing gauge
terms for two-form gauge field models \cite{gaugefix}. As we see below,
this vector ghost also exhibits a gauge invariance which, therefore,
need to be fixed. This leads to one more ghost field associated to this
subsidiary gauge invariance. Next, we also need to fix the second gauge
invariance associated to the transformation (\ref{gauge2}) and to add
its corresponding ghost field. Therefore, three constants are needed to
completely fix the gauge freedom \cite{gaugefix}. This process leads to
the following relevant additional terms that define the gauge-fixing
term in the partition function,

\begin{eqnarray}
S_{\mathrm{GF}} &=&\int d^4x\left\{ -\frac{1}{2\theta }\left( \partial^\mu
W_{\mu \nu }+\partial _\nu \psi +u\mu _WB_\nu \right) ^2+i\overline{\zeta }
^\nu \left[ \left( \partial ^2+u\mu _W^2\right) \zeta _\nu -\partial _\nu
\partial ^\mu \zeta _\mu +\partial _\nu \vartheta +u\mu _W\partial _\nu
c\right] \right.  \nonumber \\
&+&\left. i\zeta ^\nu \left( \partial _\nu \overline{\vartheta }-\mu
_W\partial _\nu \overline{c}\right) +\overline{\sigma }\partial ^2\sigma -i
\overline{c}\partial ^2c+\frac 1{2\xi }\left( \partial _\mu B^\mu \right)
^2\right\} \;,  \label{SGF}
\end{eqnarray}

\noindent
where $\psi ,\overline{c},c,\overline{\sigma },\sigma ,\overline{\vartheta },
\vartheta $ are the ghost fields and $\theta ,u$ and $\xi $ are the gauge
parameters.

We can easily perform the functional integrals over the ghost fields
appearing in Eq. (\ref{SGF}). Besides an overall normalization factor
independent of the action fields (and the background Higgs field) we get
for the quantum partition function

\begin{eqnarray}
Z &=&N\int \mathcal{D}W_{\mu \nu }\,\mathcal{D}\rho \,\mathcal{D}B_\mu \,
\mathcal{D}\overline{\eta }\,\mathcal{D}\eta \,\exp \left\{ -\int d^4x\left[
\frac{{\mu _W}^2}{2e^2\rho ^2}V_\mu ^2+\frac 14\left( \mu _WW_{\mu \nu
}-\partial _\mu B_\nu +\partial _\nu B_\mu \right) ^2\right. \right.
\nonumber \\
&+&\left. \left. \frac 12\left( \partial _\mu \rho \right) ^2-
\frac{m_\phi ^2}2\rho ^2+\frac \lambda {4!}\rho ^4-
\overline{\eta }\rho ^{-3}\eta -\frac
1{2\theta }\left( \partial ^\mu W_{\mu \nu }\right) ^2+\frac u{2\theta }
\mu_W W_{\mu \nu }\left( \partial ^\mu B^\nu -\partial ^\nu B^\mu \right)
+\frac{1}{2\xi }\left( \partial _\mu B^\mu \right) ^2\right] \right\} .
\label{Zgaugefix}
\end{eqnarray}

\noindent 
where $\overline{\eta }$, $\eta $ are the ghost fields
used to exponentiate the Jacobian $\rho ^{-3}$ in the functional
integration measure in Eq. (\ref{ZWBrho}).

Let us now compute, for instance, the effective potential for a constant
background field $\rho _c$ from (\ref{Zgaugefix}). The effective
potential for $\rho_c$ is defined as usual, by writing $\rho$ in terms
of the constant background field plus the quantum fluctuations around
this constant field configuration, $\rho =\rho_c+\rho ^{\prime }$, and
performing the functional integration over $\rho^\prime$ and remaining
fields. In the usual derivation \cite{dolan}, the effective potential
for interacting field theories is evaluated perturbatively as an
expansion in loops, which is equivalent to an expansion in powers of
$\hbar $ \cite{col-wein}. The one-loop approximation for
$V_{\mathrm{eff}}(\rho _c)$ is then equivalent to incorporating the
first quantum corrections to the classical potential $V(\rho _c)$. {}For
a general case of $N$-particle species interacting with the Higgs field,
its one-loop effective potential can be written in the generic form (in
Minkowski spacetime)

\begin{equation}
V_{\mathrm{eff}}^{\mathrm{1-loop}}(\rho _c)=V(\rho _c)\mp 
\frac{1}{2}i \sum_{j=1}^Ng_j \int \frac{d^4k}{(2\pi )^4}
\ln \left[ k^2-M_j^2(\rho
_c)\right] \;,
\label{Veff}
\end{equation}

\noindent 
where the negative sign in Eq. (\ref{Veff}) stands for
boson fields, while the positive one is for fermion (and ghost)
fields. $g_j$ labels the number of degrees of freedom for the
particle species coupled to the scalar Higgs field and $M_j(\rho
_c)$ their mass spectrum. The momentum integrals in Eq.
(\ref{Veff}), when working in the Matsubara formalism of finite
temperature field theory (see e.g. \cite{dolan,kapusta,bellac}),
are expressed as

\[
\int \frac{d^4k}{(2\pi )^4} = i\frac 1\beta \sum_{\omega _n}\int
\frac{d^3k}{(2\pi )^3}  \;,
\]

\noindent
and the four-momentum $k_\mu =({\bf k},i \omega _n)$,
where $\omega _n=2\pi nT$, $n=0,\pm 1, \ldots$, represent the
Matsubara frequencies for bosons, while for fermions we have
$\omega _n=(2n+1)\pi T$.

Using Eq. (\ref{Veff}) and from Eq. (\ref{Zgaugefix}), we obtain quantum
correction coming from the $\rho^{\prime },W_{\mu \nu },B_\mu ,
\bar{\eta},\eta $ fields. At the one-loop level, we then obtain the
effective potential for the dual Abelian Higgs model,

\begin{eqnarray}
V_{\mathrm{eff}}(\rho _c) &=&\frac{m_\phi ^2}2\rho _c^2+\frac \lambda
{4!}\rho _c^4-\frac 12i\int \frac{d^4k}{(2\pi )^4}\ln \mathrm{\det }\left[
iD^{-1}(k)\right] _{\rho ^{\prime }}-\frac 12i\int \frac{d^4k}{(2\pi )^4}\ln
\mathrm{\det }\left[ iD^{-1}(k)\right] _{B_\mu ,W_{\mu \nu }}  \nonumber \\
&-&3i\int \frac{d^4k}{(2\pi )^4}\ln \rho _c+(\mathrm{terms\;independent\;of\;%
}\rho _c)\;,
\label{Veffdual}
\end{eqnarray}

\noindent
where $[iD^{-1}(k)]_{\rho ^{\prime }}$ comes from the quadratic term in
$\rho ^{\prime }$ of the Lagrangian density, given in momentum space by

\begin{equation}
\left[ iD^{-1}(k)\right]_{\rho ^{\prime }}=
k^2 +m_\phi ^2-\lambda \rho _c^2/2\;,
\label{Drho}
\end{equation}

\noindent
while $\left[ iD^{-1}(k)\right] _{B_\mu ,W_{\mu \nu }}$ is the matrix of
quadratic terms in the gauge field $B_\mu $ and antisymmetric field
$W_{\mu \nu }$,

\begin{equation}
\left[ iD^{-1}(k) \right]_{B_\mu ,W_{\alpha \beta }}=\left(
\begin{array}{cc}
-g^{\mu \nu }k^2+(1-1/\xi )k^\mu k^\nu & -i\left( \mu _W-\frac u\theta
\right) k^\lambda g^{\mu \rho } \\
i\left( \mu _W-\frac u\theta \right) k^\alpha g^{\beta \nu } &
\mu_W^2\left( \frac{k^2}{e^2\rho _c^2}-1\right) G^{\alpha \lambda \beta
\rho }+\left( \frac 1\theta -\frac{\mu _W^2}{e^2\rho _c^2}\right)
K^{\alpha \lambda \beta \rho }
\end{array}
\right) \;.
\label{DBW}
\end{equation}
where we have used the notation

\begin{equation}
G^{\alpha \lambda \beta \rho }=\frac 14\left( g^{\alpha \lambda }
g^{\beta
\rho }-g^{\alpha \rho }g^{\beta \lambda }\right) \;,  \label{G4}
\end{equation}
and

\begin{equation}
K^{\alpha \lambda \beta \rho }=\frac 12\left( k^\alpha k^\lambda
g^{\beta
\rho }-k^\alpha k^\rho g^{\beta \lambda }\right) \;.  \label{K4}
\end{equation}

\noindent
The explicit computation of (\ref{Veffdual}) is a tedious one, but it
can be shown that all gauge dependence factorize from (\ref{Veffdual})
as terms independent of the background field and consequently can be
dropped out. {}For the generating function (\ref{Zgaugefix}) this has
been shown by the authors of the first reference in \cite{gaugefix}.
{}For the computation of the effective potential this is most easily
shown in the case of the original model. As we have emphasized before,
the model described by Eq. (\ref{Zgaugefix}) is just the dual of the
Abelian Higgs model in the covariant gauge in the polar representation
for the complex Higgs field. As such, they are physically equivalent and
the effective potential for the shifted action in (\ref{Zgaugefix}) must
lead to the same effective potential as that obtained from the original
Abelian Higgs model in the covariant gauge. This is easily seen from Eq.
(\ref{SArhochi}), where, by taking a covariant gauge fixing term, one
has the Lagrangian density

\begin{equation}
\mathcal{L}=-\frac{1}{4}
{}F_{\mu \nu } {}F^{\mu \nu }+\frac{1}{2}\left( \partial _\mu
\rho \right)^2+\frac{1}{2}\rho^2\left( \partial_\mu \chi -
eA_\mu \right)^2 + \frac{m_\phi^2}{2}\rho^2-\frac{\lambda}{4!}\rho^4
-\frac{1}{2\xi } (\partial_\mu A^\mu )^2+\bar{\eta}\rho \eta +
\bar{c}\partial^2c\;,
\label{Lcov}
\end{equation}

\noindent where $\bar{\eta},\eta $ are the ghost fields for the
Jacobian factor in Eq. (\ref{ZArhochi}) and $\bar{c},c$ are the
ghosts due to the gauge-fixing term. The effective potential for a
constant field background $\rho _c$ is defined in the usual way,
as said above. We obtain, for instance, the one-loop effective
potential,

\begin{eqnarray}
V_{\mathrm{eff}}(\rho _c) &=&\frac{m_\phi^2}2\rho _c^2+\frac \lambda
{4!}\rho_c^4-\frac 12i\int \frac{d^4k}{(2\pi )^4}\ln \mathrm{\det }\left[
iD^{-1}(k)\right]_{\rho^{\prime }}-\frac{1}{2}i\int \frac{d^4k}{(2\pi )^4}\ln
\mathrm{\det }\left[ iD^{-1}(k)\right]_{\chi ,A_\mu }  \nonumber \\
&+&i\int \frac{d^4k}{(2\pi )^4}\ln \rho_c+ i \int \frac{d^4k}{(2\pi )^4}\ln
k^2\;,
\label{Veffcov}
\end{eqnarray}

\noindent
where the last two terms in (\ref{Veffcov}) come from the functional
integration over the ghost terms of (\ref{Lcov}).
$[iD^{-1}(k)]_{\rho^{\prime }}$ is the same as before, given by
Eq. (\ref{Drho}), while
$[iD^{-1}(k)]_{\chi ,A_\mu }$ is the matrix of quadratic terms (in
momentum space) for the $\chi $ and $A_\mu $ fields,

\begin{equation}
\left[ iD^{-1}(k) \right]_{\chi ,A_\mu }=\left(
\begin{array}{cc}
\rho _c^2k^2 & ie\rho _c^2k^\nu \\
-ie\rho _c^2k^\mu & -g^{\mu \nu }(k^2-e^2\rho _c^2)+(1-1/\xi )k^\mu k^\nu
\end{array}
\right) \;.
\label{DchiA}
\end{equation}

Substituting (\ref{Drho}) and (\ref{DchiA}) in (\ref{Veffcov}), we obtain
the result

\begin{eqnarray}
V_{\mathrm{eff}}(\rho _c) &=&\frac{m_\phi ^2}2\rho_c^2+\frac \lambda
{4!}\rho_c^4-\frac 12i\int \frac{d^4k}{(2\pi )^4}\ln (k^2-M_H^2)-\frac
12i\int \frac{d^4k}{(2\pi )^4}\ln \left[ -\frac 1\xi
(k^2-M_A^2)^{3/2}k^4\rho _c^2\right]  \nonumber \\
&+&i\int \frac{d^4k}{(2\pi )^4}\ln \rho_c+i\int \frac{d^4k}{(2\pi )^4}\ln
k^2\;,
\label{Veffcov2}
\end{eqnarray}

\noindent
where $M_H^2 = - m_\phi^2 + \lambda \rho_c^2/2$ and $M_A^2 = e^2
\rho_c^2$ are the Higgs and gauge field (squared) masses as usual.
{}From Eq. (\ref{Veffcov2}) we readily see that the contributions from
the ghost fields, including the divergent contribution due to the
Jacobian coming from the radial parametrization for the scalar field
$\phi$, cancel with identical terms coming from the gauge and scalar
phase field matrix quadratic term, Eq. (\ref{DchiA}). These same
cancellations happens when working with the analogous expression for the
effective potential, Eq. (\ref{Veffdual}), in terms of the dual $B_\mu$
and $W_{\mu \nu}$ fields, including again the cancellation of the
divergent Jacobian due to an analogous contribution appearing in the
$W_{\mu \nu}$ field quadratic term, as seen from the matrix of quadratic
terms, Eq. (\ref{DBW}). All gauge dependence (on $\xi $) can be
separated from (\ref{Veffcov2}) as a background independent term that
can be dropped out. The emerging result is identical to the effective
potential obtained, e.g., in Ref. \cite{tye}.

Once the equivalence of the original and the dual model is checked and
the gauge-fixing peculiarities of the dual model can be dealt with
conveniently, we can move on to consider the contribution of singular
field-configurations with non-trivial vorticity to the effective
potential.

\section{The effective potential in the presence of
vortex--string vacuum configurations}
\label{effpot}

Let us now reinstate the contribution due to nontrivial singular
structures of the Higgs phase in the calculations of the one-loop
effective potential. This is given by the last term in Eq.
(\ref{SWBrho}), for the coupling of the antisymmetric field
$W_{\mu\nu}$ with the vorticity term due to the singular phase of
the Higgs field. As we saw, it is associated to the existence of
vortex-like solutions for the equations of motion of the action
(\ref{actini}) \cite{nielsen-olesen}. These can be associated to
string-like topological defect configurations that are either
infinite in length or forming finite-size closed loops. By open
configurations we mean the existence of magnetic monopoles at the
end points \cite{review} and we will not consider these kind of
structures here since we restrict our study only to the Abelian
theory. Also, we will only consider here field configurations
which generate closed magnetic vortex lines in the three spatial
Euclidean dimensions, since these are more suitable to the field
theoretical analysis we will adopt in the following and are also 
expected to be the dominant topology for strings close to the 
transition point \cite{review}.

The coupling term of the antisymmetric field with the vorticity source
$\omega _{\mu \nu }$, defined in Eq. (\ref{omega-munu}), is
non-vanishing for the singular term $\chi _{\mathrm{sing}}$ of the Higgs
field phase and hence this interaction term will contribute to the
action, along with the world sheet of the string. In the zero
temperature case, the source $\omega_{\mu\nu}$ is associated to the
surface element of a (tube-like) world sheet of a closed vortex-string
\cite{klee, orland, seo-suga}. {}Following the Dirac construction
\cite{dirac}, it is given by

\begin{eqnarray}
\omega_{\mu \nu }(x)=n\int_Sd\sigma_{\mu \nu }(x)\delta ^4[x-y(\xi )]\;,
\label{vort2}
\end{eqnarray}

\noindent
where $n$ is a topological quantum number, the winding number, which we here
restrict to the lowest values, $n=\pm 1$, corresponding to the energetically
dominant configurations. The element of area on the world sheet swept by
the string is given by

\begin{eqnarray}
d\sigma_{\mu \nu }(x)=\left( \frac{\partial x_\mu }{\partial \xi ^0}
\frac{\partial x_\nu }{\partial \xi^1}-\frac{\partial x_\mu }{\partial \xi^1}
\frac{\partial x_\nu }{\partial \xi^0}\right) d^2\xi
\label{sigma}
\end{eqnarray}

\noindent
and $y_\mu (\xi )$ represents a point on the world sheet $S$ of the
vortex-string, with internal coordinates $\xi^0$ and $\xi^1$. As usual,
we consider that $\xi^1$ is a periodic variable, since we work with closed
strings, whereas $\xi^0$ will be proportional to the time variable (at zero
temperature), in such a way that $\xi^1$ parameterizes a closed string at a
given instant $\xi^0$. Using (\ref{vort2}), the interaction of the string
with the antisymmetric field in the action becomes

\begin{equation}
\int d^4x\;i\pi \frac{\mu_W}eW_{\mu \nu }(x)\omega_{\mu \nu }(x)=
\frac{i}{2}\int_S d\sigma^{\mu \nu }(y)\frac{2\pi \mu_W} e W_{\mu \nu }(y).
\label{interaction}
\end{equation}

\noindent
To proceed further with the evaluation of the string contribution to the
partition function we will now introduce a (nonlocal) field associated
to the string. {}For this we take the standard Marshall--Ramond
procedure \cite{nambu2, mar-ramond} of quantizing the vortex--strings as
nonlocal objects and associate to them a wave function $\Psi [C]$, a
functional field, where $C$ is the closed vortex--string curve in
Euclidean space-time. In the second-quantized form this means that the
quanta associated to the field $\Psi$ are the vortex--strings in the
system. In introducing the vortex--string field, we first note that the
interaction term Eq. (\ref{interaction}) is in the form of a current
coupled to the antisymmetric field. Second, the coupling of the field
$\Psi [C]$ with $W_{\mu \nu }$ should respect the gauge symmetries of
the model, in particular the hypergauge one, Eq. (\ref{gauge1}). This is
fulfilled by defining the following covariant derivative term, as
proposed by Nambu \cite{nambu2},

\begin{equation}
D_{\sigma ^{\mu \nu }}(x)=\frac \delta {\delta \sigma^{\mu \nu }(x)}-
i\frac{2\pi \mu _W}{e} W_{\mu \nu }(x)\;.
\label{covar}
\end{equation}

\noindent
Here $\delta \sigma^{\mu\nu }(x)$ is to be considered as an
infinitesimal rectangular deformation of area $\delta A$ of the original
curve $A$ at a point $x$ and so the functional derivative of the string
field can be defined as the difference between $\Psi [C+\delta \sigma ]$
and the original configuration $\Psi[C]$, divided by the infinitesimal
area, taking the limit $\delta A\rightarrow 0$ (see for instance,
Refs. \cite{rey,kawai,seothesis}). The hypergauge transformation
(\ref{gauge1}) is now supplemented by the vortex--string field
transformation

\begin{equation}
\Psi [C]\to \exp \left[ -i\frac{2\pi \mu_W}e\oint dx^\mu \xi _\mu
(x)\right] \Psi [C]\;.
\label{Psigauge}
\end{equation}

\noindent
This gives sense to Eq. (\ref{covar}) as a covariant derivative, since it
commutes with the above phase change of $\Psi [C]$.

{}From the definition of the covariant derivative (\ref{covar}) the
invariant action for the string under the combined transformations
(\ref{gauge1}) and (\ref{Psigauge}) becomes (see the Appendix for 
more details)

\begin{equation}
S_{\mathrm{string}}(\Psi [C],W_{\mu \nu })=\oint_Cdx_\nu \left[
|D_{\sigma^{\mu \nu }}\Psi [C]|^2-M_0^2|\Psi [C]|^2\right] \;,
\label{Sstring}
\end{equation}

\noindent
whose explicit form and derivation has been given originally by the
authors of Refs. \cite{seo-suga, kawai} when considering the existence of $N$
connected vortex world surfaces in Euclidean space-time. The mass term for
the string field in (\ref{Sstring}) is given by  Eq. (\ref{M0square}) below.
It is  also possible to  write an  action  over local fields by defining a
functional

\begin{equation}
\hat{\psi}_C\equiv 4\left( \frac{2\pi }e\right) ^2\sum_{C_{x,t}}
\frac{1}{a^3l}\left| \Psi [C]\right| ^2,
\label{local psic}
\end{equation}

\noindent 
where $l$ is the length of a curve $C$, and $C_{x,t}$
represents a curve passing through a point $x$ in a fixed
direction $t$; also, the parameter $a$ is to be considered as a small quantity
(the lattice spacing in Ref. \cite{seo-suga}), which we choose to
be proportional to $\Lambda^{-1}$. The vacuum expectation value of
$\hat{\psi}_C$ is denoted by $\psi _C$, which represents the sum
of existence probabilities of vortices in $C_{x,t}$. In terms of
$\hat{\psi}_C$, it can be shown that the contribution of the
vortices to the quantum partition function, indicated by the last
term in Eq. (\ref{SWBrho}) and involved with the integration over
$\chi_{\mathrm{sing}}$, can be written as \cite{seo-suga}

\begin{equation}
\int \mathcal{D}\Psi [C]\mathcal{D}\Psi ^{*}[C]\exp \left\{ -\int d^4x\left[
\frac 14\left( \frac e{2\pi }\right) ^2M_0^4\hat{\psi}_C+\frac{\mu _W^2}{4}
W_{\mu \nu }^2\hat{\psi}_C\right] \right\} ,
\label{vortexcontr}
\end{equation}
where

\begin{equation}
M_0^4\equiv \frac{1}{a^4} \left(e^{\tau_s a^2}-6\right)
\label{M0square}
\end{equation}

\noindent 
and $\tau_s$ is the string tension (the total energy per unit length of
the vortex-string) \cite{kawai, seothesis}. In terms of the parameters
of the Abelian Higgs model the string tension is given by
\cite{hindmarsh} $\tau_s = \pi \rho_c^2 \, \epsilon (\lambda/e^2)$,
where $\epsilon(\lambda/e^2)$ is a function that increases monotonically
with the ratio of coupling constants. We should also note that the
factor $a^4$ in Eq. (\ref{M0square}) does not have a direct relation
with a four-dimensional space-time. Thus, the relation between $M_0$ and
$\Lambda$ ($\sim a^{-1}$) is still expected at finite temperature.

Eq. (\ref{vortexcontr}) implies, together with Eq. (\ref{SWBrho}),
that an immediate consequence of $\psi_C\neq 0$ is the increase of
the $W_{\mu \nu}$ mass. This is directly associated with a shift
in the mass of the original gauge field in the broken phase,
$M_A=e\rho_c$, as

\begin{equation}
M_A^2\rightarrow M_A^2(1+\psi_C).
\label{m-W}
\end{equation}

\noindent
Since the field $\psi_C$, defined by Eq. (\ref{local psic}), works just
like a local field for the vortex-strings, we are allowed to define an
effective potential for its vacuum expectation value $\psi_C$ in just
the same way as we do for a constant Higgs field. Since this
vortex-string field only couples directly to $W_{\mu \nu}$, at the
one-loop level the effective potential for $\psi_C$ will only involve
internal propagators of the antisymmetric tensor field. This effective
potential, at one-loop order and at $T=0$, was actually computed in Ref.
\cite{seo-suga} in the Landau gauge for the antisymmetric tensor field
propagator and it is given by (in Euclidean momentum space and at finite
temperatures)

\begin{equation}
V_{\mathrm{eff}}^{\text{1-loop}}(\psi_C)=\left( \frac e{2\pi }\right)^2
M_0^4 \psi_C+\frac{3}{2} \frac{1}{\beta} \sum_{n=-\infty}^{+\infty}
\int \frac{d^3 k}{(2\pi )^3}\ln \left[
\frac{\omega_n^2 + {\bf k}^2+M_A^2(1+\psi_C)}{\omega_n^2 + 
{\bf k}^2+M_A^2}\right] .
\label{VeffT0}
\end{equation}

\noindent
By performing the sum over the Matsubara frequencies in (\ref{VeffT0}), 
we obtain the finite-temperature expression for
$V_{\mathrm{eff}}^{\rm 1-loop}(\psi_C)$. This is a standard calculation
that gives

\begin{equation}
V_{\mathrm{eff}}^{(\beta )}(\psi_C)=\left( \frac{e}{2\pi }\right)^2
M_0^4 \psi_C + \frac{3}{2}\int \frac{d^3k}{(2\pi)^3} \omega_{\psi_C}
({\bf k})+ 3 \, \frac{1}{\beta} \int \frac{d^3k}{(2\pi)^3}
\ln \left\{ 1-\exp \left[-\beta \omega_{\psi_C}({\bf k})\right]\right\} ,
\label{VeffT}
\end{equation}
where

\begin{equation}
\omega_{\psi_C}^2(\mathbf{k})={\bf k}^2+M_A^2 \left( 1+\psi_C\right)\;,
\end{equation}

\noindent
and in Eq. (\ref{VeffT}) we have neglected the terms independent of
$\psi_C$. Eq. (\ref{VeffT}) can now be used to estimate the critical
temperature for which vortex-strings condense exactly like when we take
the effective potential for a constant scalar field to determine the
critical temperature of phase transition \cite{dolan}. By expanding
$V_{\text{eff}}^{(\beta )}$ in the high-temperature limit
$M_A\sqrt{1+\psi_C}/T \ll 1$ (this entails expanding the
temperature-dependent term in (\ref{VeffT}) just the same way we expand
the corresponding term in the usual effective potential for a constant
scalar field \cite{dolan,kapusta}), we obtain

\begin{eqnarray}
V_{\text{eff},\text{string}}^{( \beta )}(\psi_C) &=&\left(
\frac{e}{2\pi }\right)^2 M_0^4 \psi_C + \frac{3}{2} \int \frac{d^3k}
{(2\pi)^3}\omega_{\psi _C}({\bf k}) - \frac{\pi^2}{30 \beta^4}+
\frac{M_A^2(1+\psi_C)}{8\beta^2}  - 
\frac{1}{4 \pi \beta} M_A^3(1+\psi_C)^{3/2} \nonumber \\
&& - \frac{3 M_A^4 (1+\psi_C)^2}{64\pi ^2}
\ln \left[ \beta^2 M_A^2 (1+\psi_C)
\right] + \frac{3c}{64\pi ^2} M_A^4 (1+\psi_C)^2 +
{\cal O}\left[ M_A^6(1+\psi_C)^3 \beta^2 \right] ,
\label{Veff-highT}
\end{eqnarray}

\noindent
where $c\simeq 5.4076$. The momentum integral appearing in the right-hand 
side of
Eq. (\ref{Veff-highT}) represents the temperature-independent part of 
the effective potential, and it can be done directly. Using the cutoff 
$\Lambda$ we obtain for that term result

\begin{eqnarray}
\frac{3}{2} \int \frac{d^3k}
{(2\pi)^3}\omega_{\psi _C}({\bf k}) &=& \frac{3\Lambda }{16\pi ^2}\left[
\Lambda ^2+M_A^2(1+\psi _C)\right] ^{3/2}-\frac{3\Lambda }{32\pi ^2}
M_A^2(1+\psi _C)\left[ \Lambda ^2+M_A^2(1+\psi _C)\right] ^{1/2}  \nonumber
\\
&&-\frac{3M_A^4}{32\pi ^2}(1+\psi _C)^2\ln \left\{ \frac{\Lambda +\left[
\Lambda ^2+M_A^2(1+\psi _C)\right] ^{1/2}}{M_A(1+\psi _C)^{1/2}}\right\} 
\;.
\label{intk}
\end{eqnarray}

Before entering in the analysis of Eq. (\ref{Veff-highT}) it is useful
to recall that the Abelian Higgs model can support either second order
or first order phase transitions. The ratio of the coupling constants
$\alpha=e^2/\lambda$, that measure the relative intensity of the
gauge coupling $e$ and the fourth power of the Higgs potential
$\lambda$, controls these two regimes. Thus, for $\alpha\ll 1$ the gauge
coupling is quite small and the phase diagram is dominated by the second
order phase transition of the pure Higgs model. On the other hand, as
$\alpha$ gets bigger, the gauge field fluctuations are more relevant
opening the possibility of inducing a first order transition. This is
evident from the result (\ref{Veff-highT}), where the gauge
field contribution to the effective potential generates already at
one-loop order a cubic term in the Higgs background field,
which in the usual effective potential for the Higgs field is the
term that leads to  a first order phase transition in the model.

The discussion above is also in parallel with the phenomenology of the
Laundau-Ginzburg theory for superconductors, where the parameter $\kappa
\sim 1/\alpha^{1/2}$ (also called the Ginzburg parameter), measuring the
ratio of the penetration depth and the coherent length, controls the
regimes called Type II and Type I superconductors. In the former
$\alpha\ll 1$ (or $\kappa > 1$), the metal-superconductor transition is
second order and the gauge fluctuations are not important, while in the
latter $\alpha \gtrsim 1$ , the gauge fluctuations could turn the
transition first order via a Coleman-Weinberg mechanism \cite{col-wein}.
In our case, the coherent length is governed by $a\sim 1/M_H$, where
$M_H$ is here the temperature dependent Higgs mass, while the
penetration depth is proportional to $1/M_A$, where $M_A$ is the
(temperature dependent) gauge field mass. Although this effect, of the
emergence of a first order phase transition, is so weak that it is not
observable in superconductors, it could play an important role in
relativistic quantum field theory (for a pedagogic discussion of this
issues see, for instance the first volume of Kleinert's books in Ref.
\cite{kleinertI-II}). 

We turn back now to the analysis of Eq. (\ref{Veff-highT}). The lattice
spacing $a= 1/\Lambda$ can be taken as the distance between strings
\cite{rivers}. Therefore, we can consider that close to the critical
point for condensation, determined by some temperature $T_s$, $a$ can
approximately be given by the string typical radius. Then, since we are
interested in the determination of a critical point, we can write (see
for example also Ref. \cite{copeland})

\begin{eqnarray}
1/a  &\sim & m_\phi \left( 1-\frac{T^2}{T_c^2} \right)^{1/2} \;.
\label{aT}
\end{eqnarray}

\noindent
If we also use that $\rho_c$ (the Higgs vacuum expectation value) can be
expressed as

\begin{eqnarray}
\rho _c &\simeq &\sqrt{\frac{6m_\phi ^2}\lambda }
\left( 1-\frac{T^2}{{T_c}^2}\right) ^{1/2}\;,
\label{rhocT}
\end{eqnarray}

\noindent 
we see that, in the deep second order regime, where $\alpha =
e^2/\lambda \ll 1$, we have $\Lambda^2\gg M_A^2 (1+\psi_C)$ and we can
expand Eq. (\ref{intk}) accordingly. Substituting this expansion back in
Eq. (\ref{Veff-highT}) and using Eq. (\ref{M0square}), we obtain the
result (neglecting $\psi_C$-independent terms and higher order terms)

\begin{equation}
V_{\text{eff},\text{string}}^{(\beta)}(\psi_C) \simeq
\left[  \frac{e^2}{4 \pi^2 a^4}  \left(e^{\tau_s a^2}-6\right) +
\frac{3 e^2 \rho_c^2}{16\pi ^2 a^2} +
\frac{e^2 \rho_c^2}{8}\, T^2\right] \psi_C 
-\frac{e^3 \rho_c^3}{4\pi} 
\left( 1+\psi_C \right)^{3/2} T -
\frac{3 e^4 \rho_c^4 \ln \left( 2\Lambda/T \right) }{32\pi^2}  
\psi_C^2 \;,
\label{Veff-highT3}
\end{equation}

\noindent
With $a$ and $\rho_c$ given by Eqs. (\ref{aT}) and (\ref{rhocT}), we can
then see that the quantum and thermal corrections in the effective
potential for strings, Eq. (\ref{Veff-highT3}), are naturally ordered in
powers of $\alpha$. Therefore, in the regime $\alpha \ll 1$ the leading
order correction to the tree-level potential in Eq. (\ref{Veff-highT3})
is linear in $\psi_C$, while the second and the third correction terms
are ${\cal O}(\alpha^{3/2})$ and ${\cal O}(\alpha^2)$, respectively.
Thus, the linear term in $\psi_C$ controls the transition in the deep
second order regime since the other terms are all subleading in
$\alpha$. Thus, near criticality, determined by some temperature $T_s$
where the linear term in Eq. (\ref{Veff-highT3}) vanishes,
$V_{\text{eff},\text{string}}^{(\beta )}(\psi_C)\sim 0$ in the $\alpha
\ll 1$ regime.

The phase transition temperature $T_s$, which is interpreted as the
temperature of transition from the normal vacuum to the state of
condensed strings, is then determined by the temperature where the 
linear term in $\psi_C$ in Eq. (\ref{Veff-highT3}) vanishes and it is 
found to be

\begin{equation}
T_s=\frac{\sqrt{2}}{\pi a^2 \rho_c}\left( 6- e^{\tau_s a^2}- 
\frac{3 a^2 \rho_c^2}{4}   \right)^{1/2}\;,
\label{Ts1}
\end{equation}

\noindent
where the rhs of Eq. (\ref{Ts1}) is evaluated at $T=T_s$. We can now
compare the result obtained for $T_s$, given by the solution of Eq.
(\ref{Ts1}), with the usual mean-field critical temperature $T_c =
\sqrt{12 m_\phi^2/(3 e^2 + 2\lambda/3)}$ \cite{dolan}, for which the
effective mass term of the Higgs field, obtained from
$V_{\mathrm{eff}}^{(\beta)} (\rho_c)$, vanishes. Using again Eqs.
(\ref{aT}) and (\ref{rhocT}), with the result $\tau_s a^2 \sim {\cal
O}(1/\lambda)$ and in the perturbative regime $e^2 \ll \lambda \ll 1$,
it follows from Eq. (\ref{Ts1}) that

\begin{equation}
\frac{T_c-T_s}{T_c} \sim {\cal O} \left( \frac{e^{-1/\lambda}}{\lambda^{2}}
\right) \left[ 1+ {\cal O}(\alpha)\right]\;,
\label{critical shift}
\end{equation}

\noindent
with next order corrections to the critical temperatures difference
being of order ${\cal O}(\alpha)$. This
result for $T_s$ allows us to identify it with the Ginzburg temperature
$T_G$ for which the contribution of the gauge field fluctuations become
important. These results are also found to be in agreement with the
calculations done by the authors in Ref. \cite{copeland}, who analyzed
an analogous problem using the partition function for strings
configurations, in the same regime of deep second order transition.

Also, in the regime where gauge fluctuations are stronger, $\alpha
=e^2/\lambda \gtrsim 1$, the second term in Eq.\ (\ref{Veff-highT3}) of
order $\alpha^{3/2}$, induces a cubic term $\rho_c^3$ to the effective
potential, favoring the appearance of a first order phase transition
instead of a second order one. This mechanism of changing a second order
phase transition into a first order one by means of gauge fluctuations
is usually referred to as the Coleman-Weinberg mechanism
\cite{col-wein}. Coleman and Weinberg analyzed this effect in the
context of a fourth dimensional Ginzburg-Landau theory, while a similar
effect in a three dimensional theory was subsequently studied in Ref.
\cite{halperin}. 

In our context, we see that the non-trivial vacuum $\psi_c\ne 0$ above
the critical temperature $T_s$ enhance the first order phase transition
by an amount $(1+\psi_c)^{3/2}$. Here, since $T_s\sim T_c$, we see that
the driven mechanism of the first order transition is a melting of
topological defects. This mechanism is very well known in condensed
matter physics (see for instance the first reference in
\cite{kleinertI-II}) and always leads to a first order phase transition
(except in two dimensions).

\section{Conclusions}
\label{conclusions}

In this paper we have considered the evaluation of the partition
function for the finite temperature Abelian Higgs model in the context
of a dualized model realization. The advantage of adopting this
procedure is that in the dual version of the model we explicitly
identify the contribution of topological defects in the action. This way
we can identify the coupling of a topological current with the matter
fields, which in the dual field model, refers to a two-form,
antisymmetric gauge field that emerges form the dualization procedure.
We also have discussed the issue of gauge invariance in the context of
the dual model and computed all gauge fixing and required ghost terms.

The importance of the procedure we here have adopted is that now we can
take into account in the functional path integration the contribution of
not only constant vacuum field fluctuations but also those nontrivial,
inhomogeneous vacuum excitations that must emerge whenever in a theory
that exhibits spontaneous symmetry breaking the associated homotopy
group differs from the identity, which then points out to the existence
(in the broken phase) of stable topological excitations. In this paper
we have considered the case of vortex-string topological excitations of
the $U(1)$ complex Higgs field gauged model.

By considering closed magnetic fluxes in $3+1$ dimensions, we have been
able to define a local order parameter associated to the quantal
vortex-string field, making then possible to define and calculate the
effective potential associated to this vortex-string field order
parameter. Evaluating the effective potential at one-loop order and at
finite temperatures we have presented an explicit formula for the
condensation temperature for vortex-strings in the system, which then
characterizes a transition point that we have shown to lie {\it below}
the mean-field critical temperature obtained just from the contributions
of the constant scalar Higgs field vacuum expectation value to the
partition function.

We have shown that in the deep second order regime $e^2/\lambda\ll 1$,
the critical temperature for vortex condensation can be associated with
the Ginzburg temperature where the gauge fluctuations become important,
in agreement with similar results, but obtained by a different method,
by the authors in Ref. \cite{copeland}. {}Further, we have been able to
show a manifestation of the Coleman-Weinberg mechanism, by means of
which the second order phase transition can turn into a first order one
through the effect due to gauge field fluctuation contributions in the
effective potential. The vortex condensation above $T_s$ is seen to
enhance the transition. Usually, it is possible to estimate the latent
heat from the cubic term in the effective potential . However, in the
high $\alpha\equiv e^2/\lambda$ regime, where this term is important it
is not simple to calculate a reliable value for the vortex condensation
$|\psi_C|$ since we have disregarded in our model vortex interactions.

The fact that $T_s<T_c$ tempts us to interpret this transition in two
steps. As we reach the temperature $T_s$ from below, we have a vortex
condensation, but without completely restoring the broken symmetry,
obtaining in this way an intermediate phase at temperatures $T < T_c$,
since we still have a nonvanishing value for the Higgs background field
$\rho_c$. As we continue rising the temperature,
we have the final melting at $T_c$. This is usually known in the
condensed matter community as a premelting process. The possibility of
having this type of mechanism is very interesting in the context of
relativistic quantum field theory, specially related with inflationary
scenarios. However, we need to be very careful with this interpretation.
The actual window $T_s< T< T_c$ is very difficult to estimate, and is
certainly very tiny in the regime $\alpha\ll 1$ as discussed above and
seen from the result Eq. (\ref{critical shift}).
A better interpretation of the problem may be possible if both $\psi_C$
and $\rho_c$, 
the vortex-string expectation value and the Higgs vacuum 
expectation value, respectively, are considered as two
independent variables in the complete effective potential $V_{\rm
eff}^{(\beta)}(\rho_c,\psi_C)$ and study the problem as a coupled
two-field system. However,
for greater $\alpha$, where this mechanism is more suitable to be
realized, it is not possible to disregard higher order terms in the
effective potential. In particular, we have not considered in our model
vortex interactions and they could be very important in this regime,
possibly changing this scenario. Nevertheless, this premelting mechanism
is a very interesting possibility signaled by our one-loop calculation
and we believe it should deserved further attention in future works.

We also hope that the method we have employed in this paper will be
useful for further investigations, in an analytical way, of the
importance of topological excitations to phase transitions in general,
not only in the case of the Abelian gauge Higgs model studied here, but
also for non-Abelian gauge Higgs models as well, where, e.g. magnetic
monopole like excitations can also be studied in the same context.

\acknowledgments

The authors would like to thank Profs. H. Kleinert and M. Chernodub for
useful comments and also for bringing to our attention additional
previous references related to the subjected of this paper. We also
thank Conselho Nacional de Desenvolvimento Cient\'{\i}fico e
Tecnol\'{o}gico (CNPq-Brazil), {}Funda{\c {c}}{\~{a}}o de Amparo
{\`{a}} Pesquisa do Estado do Rio de Janeiro (FAPERJ), CAPES and
SR2-UERJ for the financial support. D.\ G.\ Barci would like to
acknowledge the Abdus Salam International Center for Theoretical Physics
(ICTP), where part of this work was realized, for the kind hospitality.


\appendix


\section{The dual formalism for topological field configurations}
\label{AppendixSugamoto}

In the formalism developed in Refs.
\cite{seo-suga,seothesis,kawai}, the torus-like world sheets of a
closed string contribute
to the partition function as a sum over the number and shapes of such world
sheets. The formalism is easier to understand when one considers first the
corresponding monopole problem, which involves a topological object of one
dimension less than the string problem, and one may proceed by analogy.

In the monopole case, one deals with a sum over the number and shapes of
closed loops. The monopole is taken as a relativistic particle in
interaction with an electromagnetic potential, for which we write the action

\begin{equation}
S[x_\mu (\tau )]=mn^2\int ds\,\frac{4\pi n}e\oint A_\mu (x)
\frac{dx^\mu }{d\tau }d\tau ,
\end{equation}

\noindent
where $m$, $e$ are the mass and charge of the monopole and $n$ its
topological number. For $N$ monopoles, each with its own topological number,
we have the functional integration

\begin{eqnarray}
&&\sum_{N=0}^\infty \frac 1{N!}\int \prod_{i=1}^N\mathcal{D}
y_\mu^{(i)}\sum_{\{n^{(i)}\}}\exp \left\{ i\sum_{i=1}^N\left[ -M\left(
n^{(i)}\right) ^2\oint ds+\frac{4\pi n}e\oint dy_\mu ^{(i)}A^\mu
(y^{(i)})\right] \right\}  \nonumber \\
&=&\exp \left[ \int \mathcal{D}y\sum_ne^{i\left( -Mn^2\oint ds+
\frac{4\pi n}{e} \oint dy_\mu A^\mu (y)\right) }\right] ,
\end{eqnarray}

\noindent
and because of this exponentiation one needs to consider only the action of
a single monopole. {}From now on, we put $n=1$ for the most favorable case.
The functional integral measure is defined through the introduction of a
hypercubic space-time lattice, with lattice spacing $a$. In this way, the
integral measure is reduced to the sum over all closed paths $C$. The first
term in the action is just the total length of a path; if there are $L$
steps of size $a$ on the lattice for the entire path, then its total length
is $aL$. The second term, the line integral of the field potential, is a
Wilson loop over the closed path. Defining as usual a link variable $A_\ell$
for each step $\ell $ on the path, we may write a lattice partition function

\begin{equation}
\sum_C e^{-MaL(C)+i\sum_{\ell \in C}\frac{4\pi}{A_\ell }}=
\sum_{L=0}^\infty
\frac{1}{L}\sum_nK(n,n;L),
\label{lpartition}
\end{equation}
where we have introduced the kernel

\begin{equation}
K(n,m;L)=\sum_{C(n\rightarrow m;L)}e^{-MaL+
i\sum_{\ell \in C}\frac{4\pi }{e}
aA_\ell },
\label{kernel}
\end{equation}

\noindent
for which it is understood that the sum is carried out over all paths that
go from site $n$ to site $m$ in $L$ steps. The $1/L$ factor on the
right-hand side of (\ref{lpartition}) is included in order to avoid double
counting.

In an analogous manner, one may construct an expression for the sum over the
number and shapes of the closed world sheets in the string problem
\cite{kawai, seo-suga, seothesis}. One starts with the Nambu--Goto action,
together with an interaction of the string with an antisymmetric
(Kalb--Ramond) field,

\begin{equation}
S\left[ x^\mu (\xi ^0,\xi ^1)\right] =-\tau_s \oint d^2\xi \sqrt{-g}+i\frac
\pi emn\oint d^2\xi \sqrt{-g}\epsilon ^{ab}\partial _ax^\mu \partial _bx^\nu
W_{\mu \nu }(x).
\label{nambuaction}
\end{equation}

\noindent
Here $x^\mu $ is a point on the world sheet described by the string as it
propagates through space-time and $g$ is the determinant of the sheet metric
tensor, given by\thinspace
$g_{ab}=\frac{\partial x^\mu }{\partial \xi ^a}
\frac{\partial x_\mu }{\partial \xi ^b}$, $a$, $b=0,1$, with $\xi ^0$ a
time-like coordinate variable on the world sheet and $\xi ^1$ a space-like
one. The factor $\tau_s $ in Eq. (\ref{nambuaction}) is identified with the
string tension. We follow
Kawai \cite{kawai}, differently from Seo and Sugamoto \cite{seo-suga}, and
keep the string dynamics term in our computations.

As in the monopole case, the sum over all numbers of world sheets also
exponentiate, so that we may write

\begin{equation}
e^Z=\exp \left\{ \int \mathcal{D}x\,e^{iS\left[ x^\mu (\xi )\right]
}\right\} .
\end{equation}
In the present case, the integration measure is again defined through the
use of a space-time lattice of spacing $a$ in all directions. The partition
function on the lattice then reads

\begin{equation}
Z=\sum_{{\rm all \; closed \;torus-like \; surfaces \;}S}
e^{-\tau_s a^2A(S)+i\frac{2\pi m}e\sum_{p\in S}a^2W_{p,n}},
\end{equation}

\noindent
where $a^2$ is the area of an elementary lattice plaquette and $A(S)$ is the
number of plaquettes on the surface $S$; $W_{p,n}$ is the gauge
(Kalb--Ramond) field relative to the plaquette $p$ at site $n$.

Proceeding with the analogy with the monopole case, we now have a kernel
relative to the tube-like surface of $A$ plaquettes, with the curves $C_1$
and $C_2$ as boundaries of $S$,

\begin{equation}
K(C_1,C_2;A)=\sum_{S(C_1,C_2;A)}e^{-\tau_s a^2A+i\frac{2\pi m}e\sum_{p\in
S}a^2W_p},
\end{equation}
so that

\begin{equation}
Z=\sum_{A=0}^\infty \frac 1A\sum_CK(C,C;A).
\end{equation}

Both the monopole and string kernels satisfy a recurrence equation, as they
should be seen respectively as the transition probability for the monopole
at site $m$ to go to site $n$ in $L$ steps or the string to evolve from
curve $C_1$ to curve $C_2$ sweeping a surface with $A$ plaquettes. In the
monopole case, the recurrence is established by stating that the probability
for the monopole to arrive at site $n$ in $L$ steps is in fact the product
of the probability for it to arrive at some nearest-neighbor site of $n$ in
$L-1$ steps and the probability of the last step. Therefore,

\begin{equation}
K(n,m;L)=\sum_{\pm \mu }K(n-a\hat{\mu},m;L-1)e^{-Ma+i\frac{4\pi }ea
A_{n-a \hat{\mu},\mu }},
\end{equation}
where $\ell =(n-a\hat{\mu},\mu )$ is the last link, on which we have the gauge
field $A_{n-a\hat{\mu},\mu }$. Likewise, in the string case,

\begin{equation}
K(C_1,C_2;A)=\sum_{\pm \mu \text{, }\mu \neq t}
K(C_{1,n,\mu},C_2;A-1)e^{-\tau_s a^2+i\frac{2\pi m}ea^2W_{n-a\hat{\mu},t\mu }},
\label{recurrence}
\end{equation}
where $C_{1,n,\mu }$ is a deformation of the curve $C_1$ in which one
eliminates the link $n,n+a\hat{t}$ for inclusion or deletion of a plaquette
of area $a^2$. Also, the sum is taken over all directions $\mu $
perpendicular to the curve ($t$ is a variable on the curve).

By going to the continuum limit ($a\rightarrow 0$), both kernels satisfy a
diffusion-like equation similar to that found by Stone and Thomas
\cite{stone},

\begin{equation}
\frac \partial {\partial \bar{L}}K(x,y;\bar{L})=\left[ \left(
\partial_\mu^x+i\frac{4\pi }eA_\mu (x)\right) ^2-m^2\right]
K(x,y;\bar{L}),
\label{kermon}
\end{equation}
with $\bar{L}=a^2Le^{-Ma}$ , $m^2=\frac 1{a^2}(e^{Ma}-8)$, and
\cite{seo-suga}

\begin{equation}
\frac \partial {\partial \bar{A}}K(C_1,C_2;\bar{A})=\left[ \left( \frac
\delta {\delta \sigma _{\mu t}}+i\frac{2\pi m}eW_{\mu t}(x)\right)
^2-M^2\right] K(C_1,C_2;\bar{A}),  
\label{kerstr}
\end{equation}
where $\bar{A}=a^4Ae^{-\tau_s a^2}$ and $M^2=\frac 1{a^4}(e^{\tau_s a^2}-6)$. In
fact, the differential operators on the right-hand side of both Eqs.
(\ref{kermon}) and (\ref{kerstr}) have the form of a squared covariant
derivative. In the first case the operator is

\begin{equation}
D_\mu =\partial _\mu +i\frac{4\pi }eA_\mu (x).
\end{equation}
On the lattice, acting on a scalar field $\phi (x)$, it is written as
\cite{rothe}

\begin{equation}
D_\mu \phi (x)=\frac{1}{a}\left( U_{x,x+a\hat{\mu}}^{-1}
\phi (x+a\hat{\mu}) -\phi (x)\right)\;,
\end{equation}
with $U_{x,x+a\hat{\mu}}=\exp \left[ ia\frac{4\pi }eA_\mu (x)\right]$
being the gauge field link variable. Its square then reads

\begin{equation}
D_\mu D_\mu \phi (x)=\frac 1{a^2}\left[ \sum_\mu
\left( U_{x,x+a\hat{\mu}}\phi (x+a\hat{\mu})+
U_{x,x-a\hat{\mu}}^{-1}\phi (x-a\hat{\mu})\right)
-8\phi (x)\right] ,
\end{equation}
so that when acting on the first argument of the kernel, we have

\begin{equation}
\left( D_\mu D_\mu \phi (x)+\frac 8{a^2}\right) K(x,y;L-1)=\frac
1{a^2}\sum_{\pm \mu }K(x-a\hat{\mu},y;L-1)e^{ia\frac{4\pi }eA_\mu (x-
a\hat{\mu})}\;,
\end{equation}
and, therefore,

\begin{equation}
\frac{K(n,m,L)-K(n,m,L-1)}{e^{-Ma}a^2}=\left[ D_\mu D_\mu -\frac
1{a^2}\left( e^{Ma}-8\right) \right] K(n,m;L-1),
\end{equation}
from which follows the given continuum equation

\begin{equation}
\frac \partial {\partial \bar{L}}K(n,m;\bar{L})=\left( D_\mu ^2-m^2\right)
K(n,m;\bar{L}),
\end{equation}
for $\bar{L}=e^{-Ma}a^2L$ and $m^2=\frac 1{a^2}(e^{Ma}-8)$, as stated.

A similar reasoning in one dimension less shows the string recurrence
relation (\ref{recurrence}) appearing as a discretized form of the second 
diffusion equation (\ref{kerstr}).

\end{document}